\documentclass[prb,twocolumn,showpacs,amsmath,amssymb]{revtex4}

\usepackage{graphicx}
\usepackage{dcolumn}
\usepackage{bm}

\begin{document}

\title{Thermodynamics and magnetic field profiles in
low-$\kappa$ type-II superconductors
}

\author{P. Miranovi\' c  and  K. Machida}
\affiliation{Department of Physics, Okayama University, 700-8530 Okayama, 
Japan}

\date{\today}

\begin{abstract}
Two-dimensional low-$\kappa$ type-II superconductors
are studied numerically within the Eilenberger equations of 
superconductivity. Depending on the Ginzburg-Landau
parameter $\kappa=\lambda/\xi$ vortex-vortex interaction
can be  attractive or purely repulsive. 
The sign of interaction is manifested as
a first (second) order phase transition from  Meissner to the
mixed state. Temperature and field dependence of the magnetic 
field distribution in low-$\kappa$ type-II superconductors
with attractive intervortex interaction is calculated. 
Theoretical results are compared to the experiment. 
\end{abstract}


\maketitle

Superconductors are divided into two groups, type-I and type-II, 
based on their behavior in external magnetic field. 
The underlying physics is especially transparent within the
Ginzburg-Landau (GL) model. 
It can be shown that the difference in thermodynamics originates from the
sign of vortex-vortex interaction, which can be attractive or repulsive.
The sign of the interaction is controlled by GL parameter $\kappa=
\lambda/\xi$, ratio of penetration depth and coherence length near $T_c$. 
In type-I superconductors ($\kappa<1/\sqrt{2}$) vortices attract each other. 
In type-II superconductors ($\kappa>1/\sqrt{2}$), vortex-vortex interaction 
is repulsive and the phase transition from Meissner to the mixed state at 
$H_{c1}$ is of the second order.

It has been verified experimentally since a long time that 
away from $T_c$ even
in type-II superconductors with $\kappa\sim 1$  interaction 
between vortices
can be attractive. 
Vortex attraction is manifested as a first order phase transition
from Meissner to the mixed state. 
This type of superconductivity we denote as 
type-IIa to distinguish it from the type-IIb superconductivity with purely 
repulsive vortex-vortex interaction.
Obviously attractive interaction in type-IIa superconductors will
turn into repulsive below some 
critical intervortex distance thus preventing
vortices to collapse into the normal state as in type-I 
superconductors.
Evidence for jump in magnetization and mixture of vortex and Meissner 
domains near $H_{c1}$ is found from many magnetic, calorimetric, 
decoration and neutron scattering measurements (experimental results are 
partially reviewed in Ref. \cite{lukyanchuk}). Here we refer just to a 
nicely performed experiment by Auer and Ullmaier \cite{auerandullmaier} 
in which all three types (type-I, IIa and IIb) of  superconductivity were 
reproduced in TaN by tuning the value of $\kappa$. 

Although low-$\kappa$ type-II superconductors are of less interest from the
point of view of potential application, attractive vortex interaction is 
still a challenging problem not fully understood. 
Most of the theoretical efforts to understand
interaction of vortices as a function of parameter 
$\kappa$ relied on the extended GL model. 
Utilization of Bogomolnyi method to treat
GL equations for $\kappa$ close
to $1/\sqrt{2}$ gave the further impetus for this research direction
\cite{lukyanchuk,mohamed}. 
However, range of applicability of GL formalism is confined to the region
near $T_c$. Therefore to calculate the full temperature dependence of 
$\kappa_c(T)$, boundary between type-IIa and type-IIb superconductivity, 
one should rather turn to the microscopic theory
of superconductivity. An approximate curve $\kappa_c(T)$ has been 
calculated by Kramer\cite{kramerzph} based on the asymptotic solution of 
Eilenberger equations of superconductivity. It is argued that the calculated 
curve is just the lower limit for the true one. Brandt 
\cite{brandtjump}
has obtained jumps at
$H_{c1}$ by variational approximation of his free energy functional. 
The only attempt to calculate numerically $\kappa_c(T)$ without 
approximations besides those immanent to numerical procedure is due to 
Klein \cite{klein}. 
In this thorough study clean and dirty isotropic 3D superconductors are
analyzed.

The purpose of this work is two fold.
First, we applied a new method to solve Eilenberger equations for
a vortex lattice. It appears to be efficient and faster than
the so-called ``explosion method''\cite{klein,ichioka}. 
Second, we extended previous numerical study of
2D superconductors\cite{ichioka1} to the low-$\kappa$ case where the
interesting attractive vortex-vortex interaction occurs.
Though the quantitative results 
will be different from the 3D case, the qualitative $(\kappa,T)$
phase diagram is quite similar. This is explicitly shown
by calculating phase diagram $(\kappa,T)$ of type-I, type-IIa and type-IIb 
superconductivity. 
We focused on the magnetic and temperature dependence
of calculated magnetic field profiles and compare the result
with experimental data.

Eilenberger version of BCS theory of superconductivity is proven
to be a good starting point for both analytical and numerical calculations.
For the isotropic gap Eilenberger equations  are
\begin{equation}
\left[\omega+\mathbf u\left(\mathbf\nabla+
\mathrm i\mathbf A\right)\right]f=\Psi g,
\end{equation}
\begin{equation}
\left[\omega-\mathbf u\left(\mathbf\nabla-
\mathrm i\mathbf A\right)\right]f^\dagger=\Psi^* g.
\end{equation}
These are supplemented by the self-consistency equations for the
gap function $\Psi$ and vector-potential $\mathbf A$
\begin{equation}
\Psi\ln{t}=2t\sum\limits_{\omega>0}\left[
\left<f\right>-\frac{\Psi}{\omega}
\right],
\end{equation}
\begin{equation}
\mathbf\nabla\times\mathbf\nabla\times\mathbf A=-
\frac{2t}{\widetilde{\kappa}^2}\;\mathrm{I}\mathrm{m}
\sum\limits_{\omega>0}\left< \mathbf u g\right>.
\label{selfa}
\end{equation}
Equations are written in following dimensionless units:
order parameter is measured in units $\pi T_{c}$, length in units
$R=v_0/(2\pi T_{c})$, $v_0$ is Fermi velocity,
magnetic field in units $H_0=\Phi_0/2\pi R^2$,
vector potential in units $A_0=\Phi_0/2\pi R$, energy in units $E_0=
(\pi T_{c})^2N(0)R^3$. Eilenberger parameter $\tilde{\kappa}$ is
\begin{equation}
\tilde{\kappa}^{-2}=2\pi N(0)\left(
\frac{\pi}{\Phi_0}
\right)^2
\frac{v_0^4}{(\pi T_{c})^2}\,.
\end{equation}
It is related to GL parameter $\kappa$ via $
\tilde{\kappa}^2=(7\zeta(3)/18)\kappa^2$ 
in 3D case and
\begin{equation}
\tilde{\kappa}^2=\frac{7\zeta(3)}{8}\kappa^2\,.
\end{equation}
in 2D case. Here $\zeta$ is Riemmann's zeta function.
Here, $\omega=t(2n+1)$ is Matsubara frequency with integer $n$, 
$t=T/T_c$ is reduced temperature, $\mathbf u$ is unit vector directed 
along Fermi velocity.
Eilenberger Green's functions $f$, $f^\dagger$ and $g$ are normalized so
that $g=\sqrt{1-ff^\dagger}$. Average over the
isotropic cylindrical Fermi surface $<\ldots>$ reduces to 
$(1/2\pi)\int\ldots d\varphi$, average over polar angle $\varphi$. 
Expression for the free energy density difference between superconducting and
normal state is given by, simplified with the help of Eilenberger and 
self-consistency equations
\begin{equation}
F=
\widetilde{\kappa}^2\overline{(\mathbf\nabla\times\mathbf A)^2}
-t\sum\limits_{\omega>0}\overline{\left<
\frac{1-g}{1+g}(\Psi^*f+\Psi f^\dagger)
\right>}.
\end{equation}
We use the following notation for spatial average
$\overline{C}=(B/2\pi)\int
_{\mathrm{cell}}CdS$, where $B$ is magnetic induction.
Apart from the extreme cases 
$H\sim H_{c1}$ and $H\sim H_{c2}$, which allow for an analytical solution,
the only tool at hand is numerical analysis.
Standard way to solve Eilenberger equations for a vortex lattice  is 
so-called ``explosion'' method (for details see Refs. \cite{klein,ichioka}).
Here, another  approach which takes the advantage of periodicity of
the vortex lattice  has been
adopted. 
The same approach is used by Brandt\cite{brandt1} to
solve numerically GL equations.
For this purpose   
it will be useful to introduce auxiliary functions $a$ and $b$ through the
following transformation \cite{schopohlandmaki}

$$
f=\frac{2a\exp(-\mathrm{i}\phi)}{1+ab},\quad
f^\dagger=\frac{2b\exp(\mathrm{i}\phi)}{1+ab},\quad
g=\frac{1-ab}{1+ab}.
$$
where $\phi$ is the phase of the order
parameter, $\Psi=|\Psi|\exp{(-\mathrm i\phi)}$.
Equations for auxiliary functions $a$ and $b$ are decoupled
and have the form of Riccati's differential equation 

\begin{equation}
\mathbf u\mathbf\nabla a=-\left(\omega+
\mathrm i\mathbf u\mathbf A_s\right)a+\frac{|\Psi|}{2}
\left(1-a^2\right),
\label{ricc1}
\end{equation}
\begin{equation}
\mathbf u\mathbf\nabla b=\left(\omega+
\mathrm i\mathbf u\mathbf A_s\right)b-\frac{|\Psi|}{2}
\left(1-b^2\right).
\label{ricc2}
\end{equation}
In new gauge vector-potential $\mathbf A_s=\mathbf A-\mathbf\nabla\phi$ 
is proportional to the superfluid velocity.
It diverges as $1/r$ at the vortex center (we put index $s$ to
denote its singular nature). 
Functions $a$ and $b$ are not independent but there is 
a relationship between them
\begin{equation}
a^*({\bm r},-{\bm u},\omega)=b({\bm r},{\bm u},\omega)\,.
\label{sym1} 
\end{equation}
Therefore it is sufficient to solve only the equation for
auxiliary function $a$. Also, we don't need to solve equation
(\ref{ricc1})
for all Fermi velocity directions because
\begin{equation}
a(-{\bm r},-{\bm u}, \omega)=a({\bm r},{\bm u},\omega).
\label{sym2}
\end{equation}
Reflection at the $y$ axis will change the direction of the magnetic field.
When combined with time reversal it leaves Hamiltonian
of the superconducting system invariant. One can show,
at least numerically, that the following 
equation holds
\begin{equation}
a(x,y,-u_x,u_y)=a^*(-y,-x,u_x,u_y).
\label{sym3}
\end{equation}
Therefore it is enough to solve Eq. (\ref{ricc1}) for
velocity directions $0<\varphi<\pi/2$ to recover $a$ and $b$
for all $\bm u$ via symmetry relations (\ref{sym1}), (\ref{sym2}),
and (\ref{sym3}).

Singular vector-potential has the 
following property
\begin{equation}
{\bm\nabla}\times{\bm A}_s={\bm  B}({\bm r})-
2\pi{\bm e}_z\sum\limits_{{\bm r}_i}
\delta({\bm r}-{\bm r}_i),
\label{singgauge}
\end{equation}
where ${\bm r}_i$ are positions of vortices and 
$\delta({\bm r})$
is 2D Dirac function.
For an ideal vortex lattice $z$-component of the magnetic field 
$\mathbf B({\bm r})={\bm\nabla}
\times{\bm A}$ and moduo of the gap 
function 
$|\Psi|$ has the periodicity of the vortex lattice.
The same is not true for $\Psi$ and 
${\bm A}$. However, superfluid velocity, 
i.e. ${\bm A}_s$,  is invariant under the translation 
from one to another vortex lattice cell. Therefore one must assume that 
auxiliary functions $a$ and $b$ are also periodic functions.
Then we expand $a$ in Fourier series
\begin{equation}
a({\bm r})=\sum\limits_{\bm Q}\underline{a}({\bm Q})
\exp{(i{\bm Q}{\bm r})},
\label{dft}
\end{equation}
where summation goes over reciprocal vortex lattice vectors
${\bm Q}$. Fourier transform
of a periodic function $f({\bm r})$ is defined as
\begin{equation}
\left[f({\bm r})\right]_{\mathrm{FT}}=
\underline{f}({\bm Q})=
\frac{B}{2\pi}\int\limits_{\mathrm{cell}}
f({\bm r})\exp{(-i{\bm Q}{\bm r})}\;d^2\mathbf r.
\label{ft}
\end{equation}
We split the local magnetic field $\mathbf B(\mathbf r)$ into constant 
part, magnetic induction $\overline{\mathbf B}$, and periodic part
$\mathbf B'(\mathbf r)$ with zero average over the vortex lattice cell.
Similarly, vector-potential is splitted into two parts

\begin{equation}
\mathbf B(\mathbf r)=\overline{\mathbf B}+
\mathbf B'(\mathbf r)=\mathbf\nabla\times\overline{\mathbf A}
+\mathbf\nabla\times\mathbf A'.
\end{equation}
We will assume
that $\mathbf A'(\mathbf r)$ is periodic with $\mathbf\nabla\cdot\mathbf 
A'=0$, so that all
uncertainty of the gauge is in $\overline{\mathbf A}$. 
Therefore, $\mathbf A'$ is periodic solution of the self-consistency
equation (\ref{selfa})
\begin{equation}
\mathbf\nabla^2\mathbf A'=
\frac{2t}{\widetilde{\kappa}^2}\;{\mathrm{Im}}
\sum\limits_{\omega>0}\left< \mathbf u g\right>.
\end{equation}
From the equation (\ref{singgauge}) 
one can construct singular vector-potential $\mathbf A_s$ through the 
periodic solution $\mathbf A'$, $\mathbf A_s=\mathbf A'-\mathbf \nabla\phi'$ where
\cite{kleininternal}
\begin{equation}
\mathbf\nabla\phi'=\sum\limits_{\mathbf r_i}
\frac{(y_i-y)\mathbf e_x+(x-x_i)\mathbf e_y}{(x-x_i)^2+(y-y_i)^2}.
\end{equation}
Sum converges very slowly and the rate of convergence can be improved 
by Ewald construction 
\begin{eqnarray}
&\mathbf\nabla\phi'=B{\displaystyle
\sum\limits_{\mathbf Q\ne 0}}
\dfrac{-Q_y\mathbf e_x+Q_x\mathbf e_y}{Q^2}
\exp{(-\gamma Q^2)}
\sin{(\mathbf Q\mathbf r)}+&\nonumber\\
&{\displaystyle\sum\limits_{\mathbf r_i}}
\dfrac{(y_i-y)\mathbf e_x+(x-x_i)\mathbf e_y}{(x-x_i)^2+(y-y_i)^2}
\exp{\left[-\dfrac{(\mathbf r-\mathbf r_i)^2}{4\gamma}\right]
}.
\end{eqnarray}
Parameter $\gamma$ is in principle arbitrary; we choose  $\gamma=\pi$ 
to assure good convergence for both sums.

\begin{figure}[t]
\begin{center}\leavevmode
\includegraphics[width=0.8\linewidth]{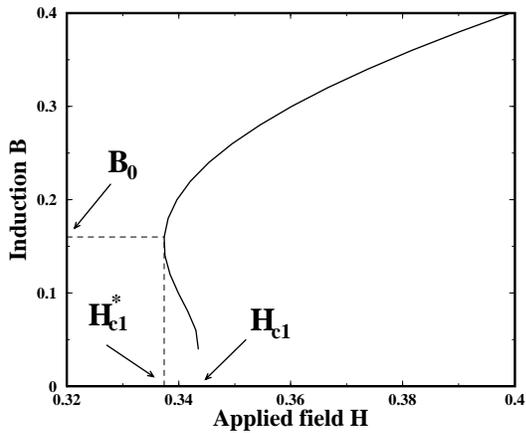}
\caption{Curve $B(H)$ for the case 
$\kappa=0.8$ and $t=0.5$.}
\label{fig2}
\end{center}\end{figure}

Self-consistency equations are solved by iteration. 
For a given vector-potential
$\bm A'$ and gap function $\Psi$, Eq. (\ref{ricc1}) is 
solved numerically.
We start the iteration procedure with ${\bm A}'=0$ and $\Psi=\rm const$.
Functions $a$ and $b$ are then used to calculate new 
values of ${\bm A}'$ and $\Psi$ with help of self-consistency equations, and
the whole process is repeated until the self-consistency is achieved.
Equation (\ref{ricc1}) is also solved by iteration.
We solved it in Fourier space where unknown are coefficients 
$\underline{a}$. Iterative procedure for unknown 
coefficients $\underline{a}$ is
\begin{equation}
\underline{a}(\mathbf Q)=\frac{\left[|\Psi|(1-a^2)/2-
\mathrm i\mathbf u\mathbf A_s a
\right]_{\mathrm{FT}}}{\omega+\mathrm i\mathbf Q\mathbf u}.
\end{equation}
Maximum summation frequency was set as $\omega_c=20t$ ($\hbar\omega_c=
20\pi k_BT_c$ in physical units). Vortex lattice unit cell is divided
into $N\times N$ mesh which turns into approximating auxiliary
functions $a$ and $b$ by sum over $N\times N$ harmonics. We set 
$N=128$. Integration over polar angle $\varphi$ in each quadrant
is performed by 30-points Gauss-Legendre integration.
Iteration procedure fails for extremely small $\omega$, i.e.
for very low temperatures
$t\le 0.05$. In all other cases it is recommended to damp the
iteration process via $a_{new}'=(c_1a_{new}+c_2a_{old})/(c_1+c_2)$
with $c_1\le c_2$.
To manipulate with eqs. (\ref{dft}) and (\ref{ft}), fast Fourier transform 
algorithm should be used.\cite{ooura}

\begin{figure}[t]
\begin{center}\leavevmode
\includegraphics[width=0.8\linewidth]{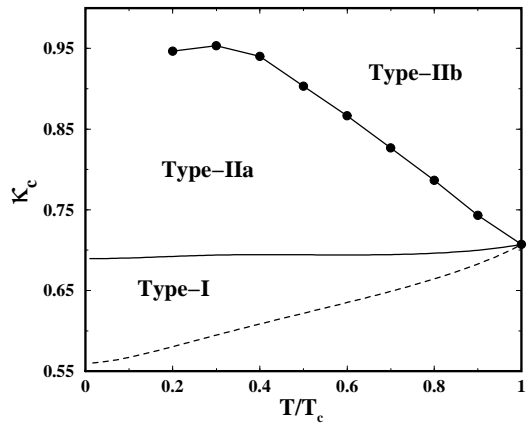}
\caption{Phase boundary between Type-I, type-IIa and Type-IIb
superconductivity. Dashed line is boundary between
type-I and type-II superconductivity in 3D case.}
\label{fig4}
\end{center}\end{figure}

Now we  concentrate on type-II superconductors with $\kappa\sim 1$. 
For a particular value of $\kappa$ and $t$
set of Eilenberger equations are solved numerically for
various values of magnetic induction $B$, so we can construct
free energy density $F(B)$. However, in the experiment we actually
control applied field $H$ and not induction $B$.  
For a given applied filed $H$ proper thermodynamic potential is Gibbs energy 
$G(\overline{\mathbf{B}},T)$ which is at minimum when the thermodynamic 
equilibrium is achieved, $\partial G/\partial B=0$. 
This condition  is rewritten via
Helmholtz free energy $F(B,T)$
\begin{equation}
H(B)=\frac{1}{2\widetilde{\kappa}^2}\frac{\partial F}
{\partial B}.
\end{equation}
If there are meta-stable states, as we expect in type-IIa superconductors,
function $B(H)$ is multi-valued and to find  a true magnetic 
induction one should look for global minimum of Gibbs energy. 
In Fig. \ref{fig2} we plotted typical example of magnetization curve
with meta-stable states, i.e. for some values of $H$ there are three
different values of magnetic induction that correspond to the extrema
of Gibbs energy. To find $B$ which corresponds to the minimum energy
one has to construct Gibbs energy density:
\begin{equation}
G=F-2\tilde{\kappa}^2 BH.
\end{equation}

For example, we found that the minimum Gibbs 
energy for $H=0.34$ correspond to the
highest value of induction $B=0.20$.
Therefore from the behavior of Gibbs energy we can reconstruct the
$B(H)$ curve. From $H=0$ up to $H=H_{c1}^*$ magnetic induction
is zero and we have Meissner state. 
Note that $H_{c1}^*$ is less than $H_{c1}=\lim\limits_{B
\rightarrow 0}H(B)$.
At $H=H_{c1}^*$ there is first order phase 
transition from the Meissner state $B=0$ to the mixed state
$B=B_0$ (jump in magnetization).  
It means that energy of well separated vortices is bigger than
vortex-lattice energy at $B=B_0$, due to attraction of vortices.
For a fixed temperature the value of induction jump $B_0$ decreases with
increasing $\kappa$ and at some critical value $\kappa_c$, $B_0=0$.
Then for all $\kappa>\kappa_c$ we have a second order phase transition
from the Meissner to mixed state, i.e. type-IIb superconductivity.

Calculations are performed for various temperatures $t=0.2-0.9$ to find
the critical parameter $\kappa_c$ that separates type-IIa and type-IIb 
superconductivity. The results are shown on Fig. \ref{fig4}. Boundary
line is qualitatively similar to the 3D case obtained by different
numerical scheme.
Boundary between type-I and type-II superconductivity can be obtained from the
condition $H_c=H_{c2}$. 
Calculation itself is trivial and we
did it merely to show the quantitative difference between 
3D (shown by dashed line in Fig. \ref{fig4})
and 2D case. Compared to 3D case, boundary between type-I and 
type-II superconductivity in 2D superconductors
is almost constant and changes only few percent down to $T=0$.
In 3D case there is possibility for cross-over from type-I to type-II
superconductivity as one goes from $T_c$ to $T=0$.

We obtained selfconsistent solution for the order parameter
$\Psi({\bm r}, t)$ and field distribution ${\bm A}({\bm r},t)$
for various $\kappa$, $B$ and temperature $T$. Only one interesting
aspect will be presented: magnetic field profile as viewed via 
nuclear magnetic resonance (NMR) and muon spin rotation
($\mu$SR). These are standard experimental techniques that probe
the magnetic field profile in the mixed state.
NMR absorption intensity in a small frequency interval
$\Delta\omega$ of the ac field is proportional to the fraction
of the VL cell area where the field value is in the corresponding interval
$\Delta\omega\propto\Delta H$, i.e. 
signal is proportional to $\int\delta(H-H({\bm r}))
\mathrm{d}{\bm r}$.
NMR line shape has the logarithmic singularity at the saddle point
of magnetic field. In $s$-wave superconductor from the relative position
of the NMR peak one can make conclusion about the VL structure.

\begin{figure}[t]
\begin{center}\leavevmode
\includegraphics[width=0.8\linewidth]{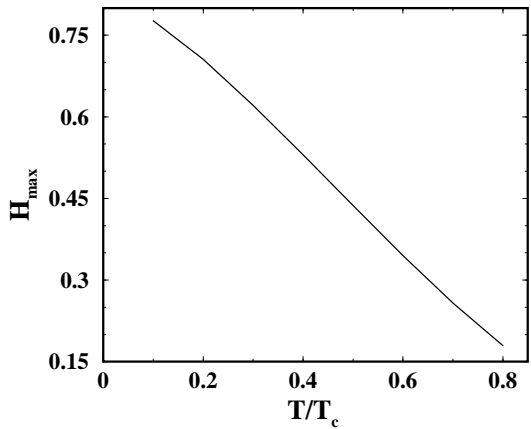}
\caption{Temperature dependence of magnetic field in 
the center of the vortex for parameters given in the text.}
\label{fig5}
\end{center}
\end{figure}

Kung \cite{kung} measured NMR profile in V foil 
as a function of temperature. The applied field was perpendicular
to the foil plane. In this geometry magnetic induction was fixed
and almost temperature independent. It was found that 
the maximum magnetic field
(field at the vortex center) is linear in $T$ down to the lowest
attainable temperature $0.1T_c$ in that experiment.
Delrieu \cite{delrieu} calculated magnetic field profile
in the limit $T=0$ and small magnetization. He found singularities 
at the positions of field maximum and minimum which have conical shape 
compared to parabolic shape near $T_c$. It was shown that 
the slope
$\mathrm{d}H_{max}/\mathrm{d}T$ is finite at $T=0$.

We calculated temperature dependence of $H_{max}$ in superconductor
with $\kappa=0.8$ for fixed magnetic induction $B=0.16$. The result
is shown in Fig. \ref{fig5}. Overall behavior $H_{max}(T)$ resemble 
the linear behavior reported in \cite{kung}. The result is in agreement 
with the calculation of Delrieu if the curve is extrapolated to $T=0$.
We note that in Ref. \cite{klein} $H_{max}$ has a more pronounced
convex curvature. Therefore, in spite the value of 
magnetic
penetration depth 
is saturated already below $0.4T_c$, magnetic field distribution 
is temperature dependent down to very low temperatures.

In summary, we presented the new method to solve selfconsistently
Eilenberger equations in the mixed state. Its
efficiency is demonstrated in  the problem of low-$\kappa$ 2D
superconductors. Being faster then previous  techniques,
we hope it will stimulate further numerical studies of
type-II superconductors.

\end{document}